# I. Introduction

Among non-thermal plasma sources, atmospheric pressure plasma jet (APPJ) devices have shown an increasing interest over the past three decades. This growing attention is primarily due to their ability to operate at atmospheric pressure, thereby eliminating the need for costly vacuum systems or low-pressure environments. Additionally, their gas temperature can be maintained at levels close to ambient air, ensuring compatibility with sensitive materials and environments [Schutze, 1998], [Chang, 2012]. These characteristics make APPJs highly versatile and suitable for a wide range of applications, including gas processing, surface modification, and, more recently, biomedical applications.

In gas processing, APPJs effectively dissociate $CO_2$ through selective electron-driven mechanisms, minimizing energy consumption. Contrarily to thermal plasmas or combustion, APPJs avoid bulk heating and thermal decomposition, leveraging non-equilibrium pathways to reduce unwanted side reactions [Jen, 2024], [Xia, 2024]. In surface modification, APPJs are employed to decontaminate complex or fragile medical devices as well as chemical and biological warfare agents [Herrmann, 1999], [Weltmann, 2012], [Zhang, 2022]. In recent years, APPJs have gained prominence in medical applications, particularly in wound healing, where they can inactivate bacteria around wounds and promote fibroblast proliferation in wound tissue [Xu, 2015], [Bekeschus, 2016]. Additionally, APPJs have been explored in oncology, such as in melanoma treatment, through *in vivo* studies on C57 mice [Mashayekh, 2015].

For medical applications, APPJ devices must adhere to strict safety standards, especially when applied to living organisms. A key requirement is the generation of cold plasma, ensuring that gas temperatures remain below 40°C to prevent irreversible cellular damage [Zenker, 2008]. Unlike APPJs, the argon plasma coagulator (APC), another plasma device widely used in clinical settings, operates at higher temperatures, ranging from 60-80°C for blood coagulation and up to 100°C for tissue ablation [Cai, 2023], [Hoffmann, 2015], [Zenker, 2008]. In addition to the thermal constraint (<40°C) for medical use, APPJs must comply with the international standard for medical electrical equipment safety, which limits patient leakage current to a maximum of 10 µA under normal operating conditions [Backles, 2006]. This condition is related both to the IEC 60601-1 standard which applies to medical electrical equipment and to the IEC 60479-1 standard which focuses on the effects of current passing through the human body. Another safety consideration is ultraviolet (UV) radiation, since excessive exposure can lead to DNA damage, erythema, or tissue necrosis. Fortunately, UV emissions from APPJs typically remain well below the permissible daily exposure limit which is estimated to 3 mJ/cm² [Nascimento, 2024]. Once these safety criteria are satisfied, APPJs can be fine-tuned for the controlled production of reactive oxygen and nitrogen species (RONS). These species are the primary agents of APPJ therapeutic effects, facilitating the modulation of cellular signaling pathways and immune responses [Privat-Maldonado, 2019], [Wende, 2018].

In previous studies, our team in partnership with biologists and an endoscopic practitioner from CRSA, Saint-Antoine Hospital (Paris, France) successfully developed and optimized APPJ and TPC (transferred plasma catheter) devices to treat cholangiocarcinoma, a solid tumor originating in the biliary tracts. The antitumor efficacy of these devices was demonstrated in immunocompromised mice, where a CCA cell line was subcutaneously grafted onto the flanks of the animals [Vaquero, 2020]. Then, to facilitate the translation of these results from bench to bedside, a feasibility study was conducted using a series of *ex vivo* porcine digestive systems, which serve as preclinical models closely mimicking human anatomy [Decauchy, 2022]. A key challenge was ensuring precise control of plasma dynamics within such confined cavities while eliminating electrical and thermal risks. To address this, cold plasma was generated in the form of guided streamers propagating at a specific, regular repetition frequency. This approach provided superior control and stability compared to traditional sinusoidal excitation, which results in random streamer distribution. By leveraging this precise control mechanism, the study demonstrated the potential for safe and effective plasma delivery within the biliary system.

The propagation of streamers (guided or branching, positive or negative) mostly depends on (i) the medium where they are generated, gas or liquid; (ii) the surface they are interacting with, dielectric or conductive; and thus the distance between the source and this surface [Njidam, 2020]. While several studies have investigated the dynamics of guided streamers propagating from APPJ devices in plasma gun configuration [Pinchuk, 2021], no similar characterization has been conducted for APPJs containing a transmission line such as in the TPC configuration. Furthermore, our study aims to address this knowledge gap in a context of clinical transfer.

In this context, one has to keep in mind that each patient has a unique impedance, which varies based on physiological factors such as hydration, skin condition, and overall health. For instance, a pregnant woman does not have the same impedance as a child, just as a person with moist hands differs from someone with dry skin [Goyal, 2022], [Morin, 2020], [Buchholz, 2004]. This variability poses a challenge in ensuring that a cold plasma jet device operates safely for all patients. A device calibrated for one individual may induce unintended effects in another with lower impedance. To mitigate these risks, two approaches are possible: designing a personalized plasma source for each patient or developing a universally safe device. We opted for the latter, ensuring that our transferred plasma catheter remains safe even in extreme conditions, such as the theoretical "zero-impedance patient," represented by a grounded metallic target. This configuration provides the most stringent scenario for charge transfer and current dissipation, allowing us to define conservative safety limits. Therefore, our study aims to establish a robust calibration framework to ensure the safe and effective application of TPC devices in clinical settings. Such study is carried out by combining spatiotemporally electric and optical measurements.

## 2. Materials & Methods

### 2.1. Experimental plasma setup

The transferred plasma catheter is depicted in **Figures 1a-1b**. It comprises a dielectric barrier chamber in quartz (150 mm in length, 3 mm of inner diameter and 1 mm in thickness) and a flexible capillary in PTFE (2 meters in length, 2 mm of inner diameter and 0.5 mm in thickness). The dielectric barrier chamber is surrounded by a high voltage electrode connected to a nanopulse power device (Nanogen 1 – RLC Electronic company) coupled with a DC high-voltage power supply (SLM 10 kV 1200 W – Spellman company). In addition, a transmission line (tin-plated and PTFE-shielded copper wire, 0.56 mm in diameter) is placed inside the flexible capillary. The tip of this transmission line is located 5 mm before the end of the capillary, as sketched in **Figure 2a**. The device is supplied in helium gas at 1 slm. The target, a grounded metal plate measuring 60 mm × 80 mm × 2 mm, is positioned at varying gap distances (from 2 mm to 18 mm) from the capillary outlet.

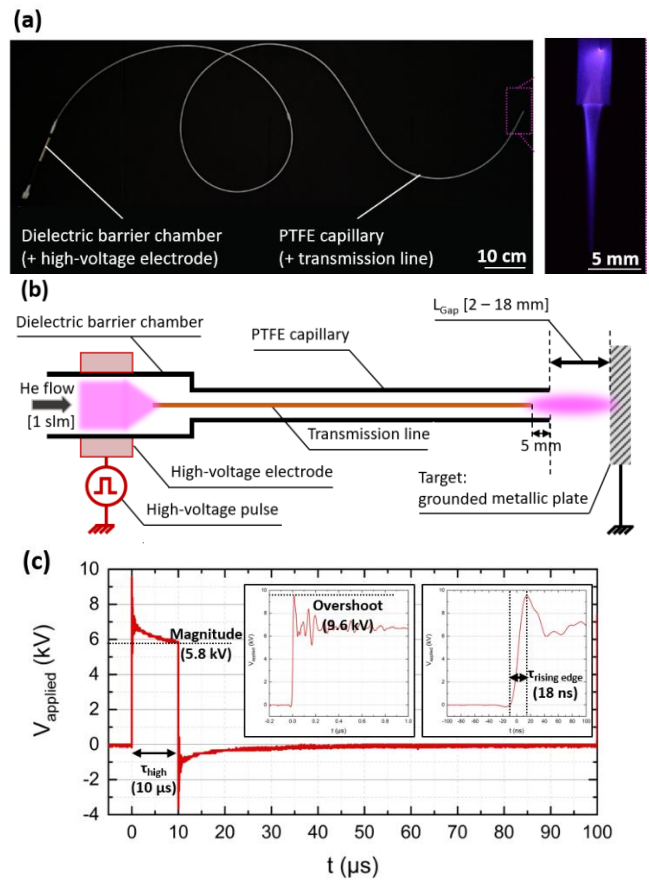

*Figure 1. Device scheme and applied voltage characteristics. (a) Photograph of the catheter with a zoom of its distal end emitting a plasma plume in free-jet configuration. (b) Transfer plasma catheter and target represented for the current conditions: helium flow rate of 1 slm, high-voltage supply of 5.8 kV, 10 kHz, 10% and $L_{Gap}$ variation from 2 to 18 mm. (c) Positive squared pulse obtained for a period of the applied voltage (T = 100 µs) and a $L_{Gap}$ of 2 mm, focused on the rising edge are added to measure the overshoot and its rising time.*

As shown in **Figure 1c**, the applied voltage waveform has a 5.8 kV amplitude with a 10 kHz repetition frequency and a 10% duty cycle ($\tau_{high}$ = 10 μs). Each pulse exhibits rising and falling times as short as 18 ns and an overshoot of 9.6 kV. In these conditions, the plasma plume emerges 17 mm in ambient air without target, representing the free-jet condition as shown in **Figure 1a**.

## 2.2. Electrical characterizations

Each guided streamer of the plasma plume can be characterized through electrical measurements using multiple probes connected to a digital oscilloscope (Wavesurfer 3054, Teledyne Lecroy), as shown in **Figure 2a**. The voltage applied to TPC ($V_{applied}$) is measured using a high-voltage probe (P6015A 1000:1, Tektronix), while the current through the target ($I_{target}$) is recorded with a current monitor (Pearson 2877, Pearson Electronics). The set-up comprises electromagnetic emission sources such as the high-voltage supply, its transformers and cables, the latter one acting as an antenna and diverse electromagnetic captors such as the BNC cable connecting the current probe to the oscilloscope, the oscilloscope port itself and the grounded metal plate. The streamer current ($I_s$) is isolated from the surrounding electromagnetic environment and capacitive components ($I_{ps}$) following the method proposed in [Decauchy*, 2022].

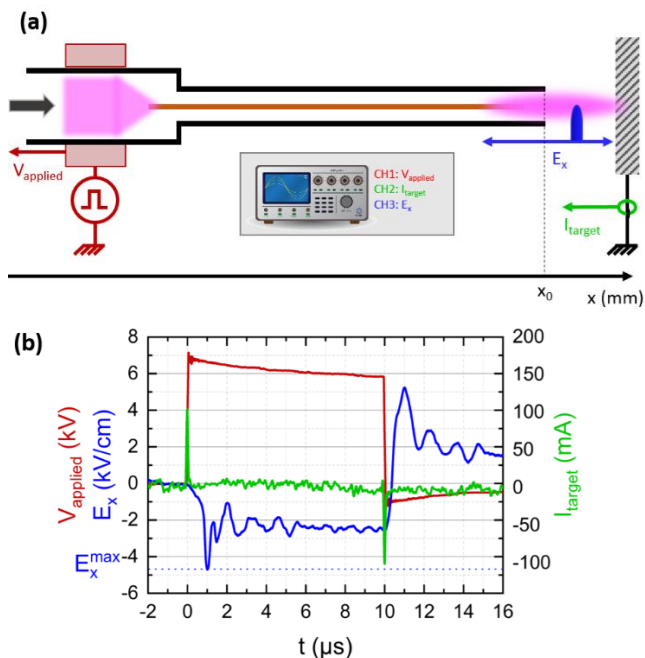

*Figure 2. Electrical characterization setup and properties. (a) Electrical probes location in the setup. (b) Applied voltage, axial electric field and target current measured for one period of the pulse for a Pockels probe located at x = 8 mm in the gap of 15 mm.*

The local electric field is measured using a Kapteos probe (eoProbe - ETX) with measurements taken along the capillary ($x \leq x_0$) and along the gap ($x \geq x_0$), where $x_0$ corresponds to the capillary outlet. This device operates based on the Pockels effect in birefringent crystals, correlating the measured voltage to the electric field ($E_x$) via an antenna factor (AF = 120.5 ± 0.2 dB/m under the experimental conditions), as provided by the optoelectronic converter. The antenna factor is quantified by the optoelectronic converter (eoSense), calibrated by the company using an adiabatic TEM cell (eoCal) and a vector network analyzer (VNA) [kapteos, 2021]. As the probe partially interacts with the guided streamers upon their propagation, it introduces a slight perturbation that may result in a marginal extension of the streamer path length. Two measurements are performed, filtered, and subtracted: the first with the entire experimental setup in action, and the second without helium supply to isolate the electromagnetic noise inherent to the plasma. The low-pass filter's cutoff frequency is again set to 10 MHz.

The temporal evolution of streamer propagation is controlled by the voltage pulse and monitored via current and electric field measurements, as shown in **Figure 2b** for a fixed position (x = 8 mm) in the gap. At the voltage rising edge, a positive current ($I_{target} \propto V_{applied}$) is generated, immediately followed by a negative electric field magnitude ($E_x = -dV_{applied}/dx$), with the opposite occurring at the falling edge.

## 2.3. Fast-ICCD imaging

The propagation of guided streamer trains is investigated using an intensified charge-coupled device (ICCD) camera (Istar DH340T, Andor) equipped with a 2048 × 512 CCD matrix of 13.5 μm × 13.5 μm pixels. The camera's intensifier provides a minimum optical gate width of less than 2 ns and a quantum efficiency exceeding 10% across the 200–550 nm wavelength range. A macro lens (AF-S Nikkor 24-85 mm, Nikon) is attached to focus on the plasma region of interest, blocking UV emissions. Temporal imaging is achieved through external synchronization of the camera, triggered by the rising edge of the applied voltage. The typical delay between the input trigger pulse and the actual ICCD triggering is 100 ns, with a jitter of 35 ps rms.

During a 5 s-exposure, the camera captures images for a fixed aperture time of 2 ns, at the same instant of each cycle. The digital storage time of the camera did not impose limitations on the acquisition process, as the ICCD is triggered on the rising edge of each high-voltage pulse, ensuring consecutive image acquisition without skipping pulses. This functionality is further supported by the Integrate-On-Chip (IOC) gating mode, which enables efficient signal accumulation while minimizing exposure time.

After the exposure period, the images are combined, resulting in a single composite image that accumulates approximately $5 \times 10^5$ frames (**Figure 3a**). This method generates a series of images representing the streamer's emissions at the same instant during successive cycles, with each image separated by a 2 ns time step.

To determine the location of the guided streamer's head ($x_{sh}$) in each image, emission intensities are summed perpendicular to the propagation direction, as illustrated in **Figure 3a**. The head corresponds to the region of highest intensity, while the tail is not fully observed, as it originates at the distal end of the transmission line. In some experimental conditions, the tail near the transmission line may appear brighter than the head, especially when the ICCD captures the entire setup. In these cases ($L_{Gap}$ > 10 mm), locating the streamer head based on the brightest region is not reliable. Therefore, the streamer head's position is defined as the first intensity maximum, starting from the target. This approach provides the streamer head's position as a function of time (**Figure 3b**).

To calculate the instantaneous velocity of the streamer's head ($v_{sh}$), the position versus time curve is fitted by a Boltzmann sigmoid function and differentiated. The result, shown in **Figure 3c** for a 16 mm-$L_{Gap}$, illustrates the velocity evolution. Additionally, the tangent velocities are estimated using a linear regression at the capillary outlet ($v_0$) and further in the gap, at the target vicinity ($v_T$) as given in example in **Figure 3b**. These values facilitate comparison between different gap conditions.

configuration utilizing two converging lenses (LA4380-UV, Thorlabs) with a focal length of 100 mm. This setup achieves a spatial resolution of 1.0 mm within the plasma. As shown in **Figure 3d**, spectral measurements are conducted at three distinct positions along the plasma propagation axis: inside the capillary (x = −10 mm), at the capillary outlet (x = 1 mm), and in the vicinity of the target (x = 9 mm). Emission collected within the capillary is negligibly attenuated by the PTFE capillary wall owing to its native transparency. This configuration allows for precise characterization of plasma-generated species and their spatial distribution along the discharge pathway.

## 3. Results & Discussion

### 3.1. Influence of the gap in the dynamics of the streamer current

The investigation of the maximum intensity of the streamer current as a function of the gap distance is crucial not only to assess the maximum current that propagates from the TPC to the target, but also to evaluate how the gap affects the streamer's ability to propagate over several centimeters. Understanding the current flowing through the conductive target is particularly important for cold plasma-assisted digestive endoscopic procedures, where the patient is grounded during plasma generation. According to TS-60479-1 standards from the International Electrotechnical Commission (IEC), the potential risks are in the following ranges in the case of a sinusoidal current: magnitudes between 100 mA and 100 A for durations between 100 μs and 10 ms, as carefully exposed in [Decauchy, 2022]. Otherwise, if the current is too low, it indicates insufficient electrical contact between the plasma and the target, limiting the effectiveness of the intended treatment. Therefore, determining the maximum gap distance at which this TPC remains effective is essential.

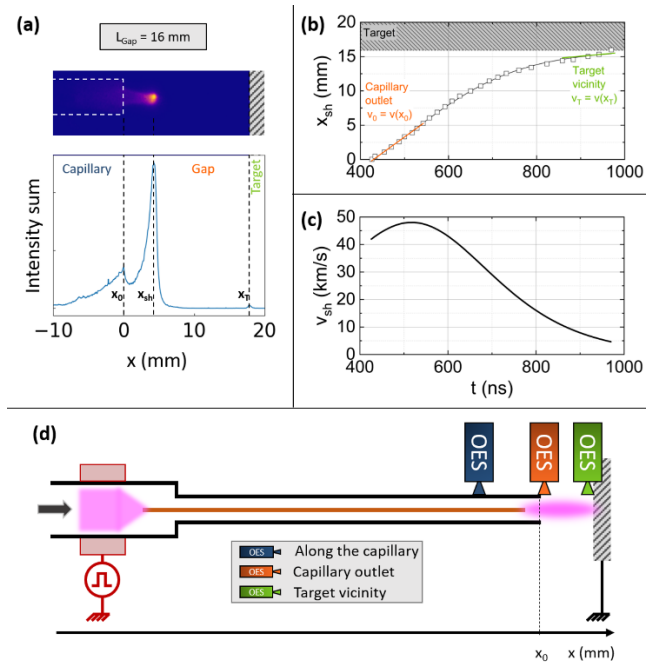

*Figure 3. Optical characterization setup and methods. (a) Picture of the streamer for $L_{Gap}$ = 16 mm, at t = 528 ns and the intensity sum collected along the x axis of the picture, specifying the outlet of the capillary $x_0$, the streamer's head location $x_{sh}$ and the target location $x_T$. (b) Position of the streamer's head during its propagation; the orange line is the tangent of the curve at the outlet of the capillary, given the tangent velocity $v_0$ at this location, respectively the tangent velocity $v_T$ at the target vicinity for the green line. (c) Instantaneous velocity of the head of the streamer, derived from the positions, fitted by a Boltzmann sigmoid law. (d) Optical emission spectroscopy setup for a $L_{Gap}$ of 10 mm.*

### 2.4. Optical Emission Spectroscopy

The plasma's emissive species are collected using an Optical Emission Spectrometer (SR-750-B1-R, Andor), which displays a 750 mm focal length and a 1200 grooves/mm grating, blazed at 500 nm, coupled to the previous ICCD camera (Istar DH340T, Andor). The camera's internal trigger is activated, accumulating the signal 3 times over 0.5 seconds. The entrance slit of the spectrograph is connected to an optic fiber (SR-OPT-8014, Leoni fiber optics) featuring a core diameter of 100 μm. To precisely observe the location along the axis of the plasma plume, we employed a 4f-

In our setup, the grounded target represents the most stringent scenario, as it disregards the patient's resistive and capacitive properties. As shown in **Figure 4a**, increasing the gap reduces the conductivity of the medium, making the streamer peak indistinguishable at gap distances greater than 10 mm. For gap distances less than 10 mm, $I_s$ is isolated by setting all preceding values to zero, and the charge carried by the streamer is calculated by integrating $I_s$ over five periods of the signal. The maximum current ($I_{max}$) decreases linearly with increasing gap distance, from approximately 100 mA at a 2 mm-$L_{Gap}$ to around 10 mA at a 10 mm-$L_{Gap}$ (**Figure 4b**). These current magnitudes are obtained from a pulse signal during less than 1 μs, ensuring a device that can be deemed safe for real patients. Furthermore, as the gap distance increases, recombination of species with ambient air reduces the number of charged particles, and additional current dissipation through pressure gradients and viscosity limit plasma propagation.

The delay between the rising edge of the applied voltage, which also corresponds to the peak of $I_{ps}$, and the maximum streamer current ($I_s$) represents the time required for the primary streamer to propagate from the tip of the transmission line to the target. **Figure 4c** shows that this delay increases exponentially with the gap distance. The transmission line's tip produces a localized "spike" effect, in contrast to the grounded target, which acts as a large, flat electrode. As the gap distance increases, the electric field near the target weakens, slowing the streamer's propagation.

The charge carried by the streamer is calculated through temporal integration of its current. **Figure 4d** demonstrates that the streamer charge decreases with increasing $L_{Gap}$, consistent with the trends observed for maximum current. The streamer charge drops from about 60 nC at a 2 mm-$L_{Gap}$ to 10 nC at a 10 mm-$L_{Gap}$. However, the method used to isolate the streamer current introduces some uncertainties in determining the end of the current profile, leading to a non-linear regression behavior as a function of gap distance.

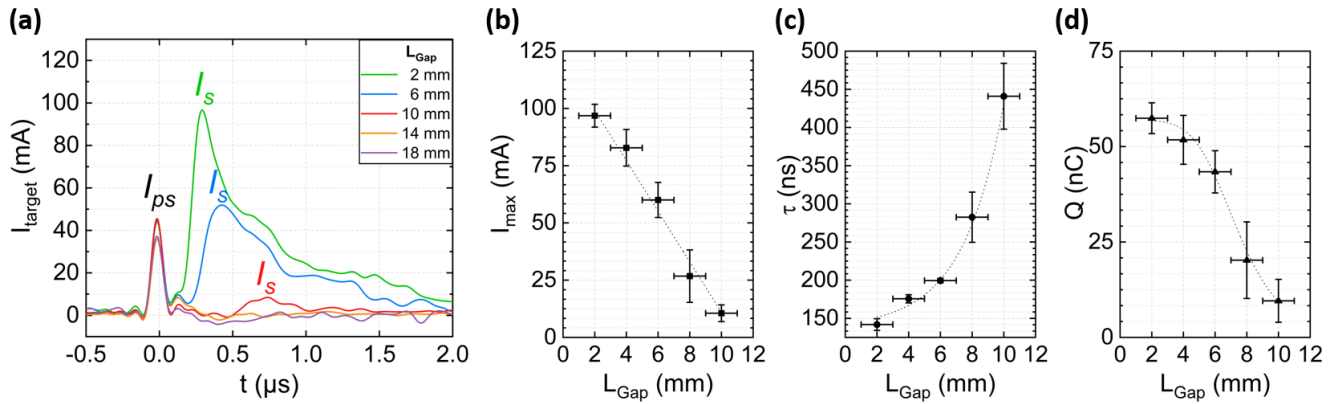

*Figure 4. Properties of streamer current dependance on the gap distance. (a) Streamer $I_s$ and pre-streamer $I_{ps}$ current peaks after numerical treatment for various gap distances from 2 to 18 mm. (b) Maximum value of the streamer current. (c) Delay between the beginning of the streamer current and the maximum of the pre-streamer current. (d) Charge carried by the streamer per period. Data are extracted from 3 replicates of a 5 period-signal, preliminarily filtered and smoothed.*

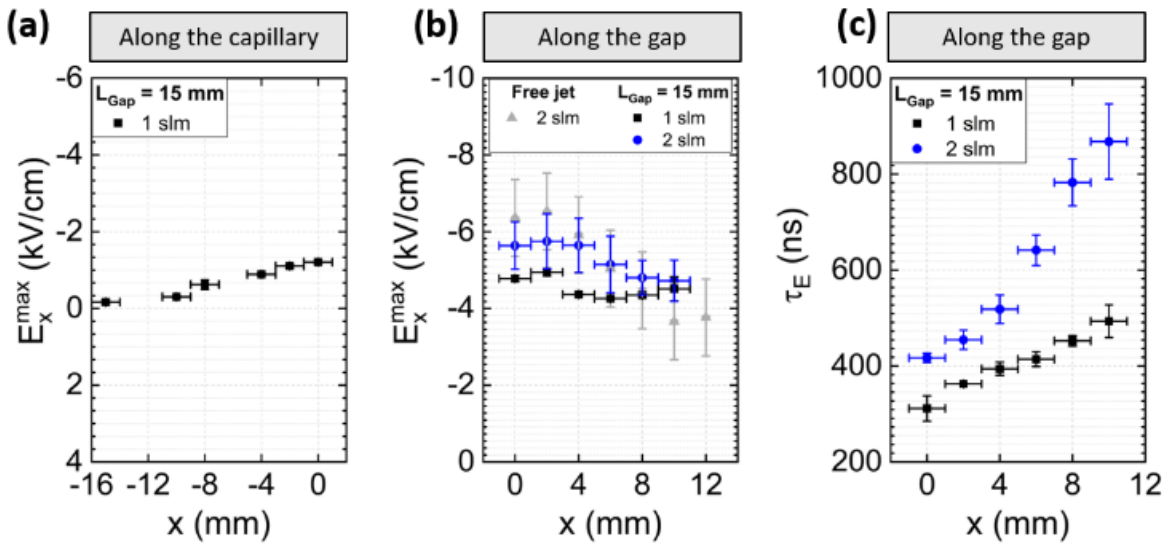

*Figure 5. Electric field properties as a function of the probe location x for $L_{Gap}$ = 15 mm. (a) Maximum magnitude of the electric field ($E_x^{max}$) along the capillary for a 1 slm-helium flow rate. (b) $E_x^{max}$ inside the gap for helium flow rates of 1 slm and 2 slm. The free jet case is depicted for 2 slm as a control. (c) For the previous conditions, the delay between the rising edge of the voltage pulse and the beginning of $E_x^{max}$.*

## 3.2. Electric field profile along the plasma axis

The electric field profile is measured locally along the plasma axis for a fixed gap of 15 mm along the capillary and within the gap area. Initially, the maximum magnitude of the electric field ($E_x^{max}$) is measured along the capillary supplied with He flow rate at 1 slm, up to 15 mm before the capillary outlet (**Figure 5a**). Since the probe cannot be positioned inside the capillary because of its dimensions, $E_{max}$ in this region is lower than the one in the gap, around 1.5 kV/cm compared to approximately 5 kV/cm in the gap (**Figure 5b**).

Subsequently, the $E_x^{max}$ is measured in the gap region under three conditions: 1 slm with a target positioned at 15 mm, 2 slm with a target positioned at 15 mm showing a more defined electric field peak, and 2 slm without a target representing a free jet reference case. **Figure 5b** presents $E_x^{max}$ under these conditions, showing that the peak of the electric field consistently ranges between 4 and 6 kV/cm. In the free jet case, the electric field decreases as the Pockels probe moves farther from the capillary outlet. Introducing a grounded target at a 15 mm-$L_{Gap}$ with the same helium flow rate does not significantly alter the peak electric field, although the rate of decline is slower, and the field tends to stabilize. A similar trend is observed for the 1 slm helium flow rate, where the $E_x^{max}$ peak remains relatively constant across all positions within the gap.

The maximum magnitude of the electric field corresponds to the position of the streamer head. The delay between the onset of this maximum and the rising edge of the voltage pulse corresponds to the time required for the streamer head to reach the probe. **Figure 5c** illustrates this delay as a function of probe location for the previously described electric field profiles. In both helium flow rate conditions, the delay increases as the probe moves further away, with a more pronounced increase for the 2 slm helium flow rate.

Beyond their relevance in assessing plasma propagation, our electric field measurements provide valuable insight into potential therapeutic applications of the transferred plasma catheter. While the primary focus in plasma medicine has been on RONS, recent studies suggest that the electric field itself plays a significant role in biological effects, particularly in tissue regeneration and wound healing [Cortazar, 2024]. Electric fields have long been recognized for their influence on cellular migration, proliferation and differentiation but also on apoptosis [Radeva, 2009], [Katoh, 2023]. Our measurements indicate that the transferred plasma catheter generates electric field magnitudes within a relevant range, suggesting a potential synergy between plasma-generated RONS and electrostimulation. This could be particularly beneficial for endoscopic applications in gastrointestinal medicine, such as the treatment of stomach and duodenal ulcers.

## 3.3. Spatio-temporal dynamics of guided streamers: contact time, counter-propagation & Residual light

Image series are acquired for gap distances ranging from 2 to 18 mm, with each image separated by a 2 ns step. During plasma generation, the streamer propagates between the tip of the transmission line and the target. **Figure 6** shows the sum of intensities along the x-axis for each gap distance, depicting the streamer's behavior as it propagates partially through the capillary and into the gap.

Regardless of the gap distance, a primary streamer consistently emerges from the transmission line tip within a few hundred nanoseconds after the rising edge of the voltage pulse. The streamer head propagates through the catheter for several nanoseconds before traveling through the air, eventually reaching the target. However, the streamer's propagation and intensity vary based on the gap distance. The time required for the primary streamer to form at the tip of the transmission line, as well as the time needed to reach the target, increases with the gap distance. This variation can be attributed to both the distance the streamer must travel and its velocity. For example, with a 4 mm-$L_{Gap}$, the streamer head reaches the target in 110 ns, while for a 16 mm-$L_{Gap}$, the streamer remains within the capillary at that same time. Notably, for gap distances greater than 10 mm, significant differences in streamer behavior are observed.

For gap distances less than 10 mm (**Figures 6a–6e**), the following phenomena are observed:
- *Primary Streamer Propagation*: The head of the primary streamer, defined by the boundary between dark and bright regions, shows increasing luminous intensity during propagation, peaking upon reaching the target. The steepening of this boundary as the streamer exits the capillary and approaches the target suggests an acceleration during propagation (mostly seen in **Figure 6e**).
- *Primary Streamer-Target Interaction*: When reaching the target, a highly intense contact point forms, persisting for a few nanoseconds corresponding to a charge accumulation in that region (mostly seen in **Figure 6d**).
- *Counter-Propagation*: Immediately after the contact, a counter-propagating ionization wave is observed, moving in the opposite direction of the primary streamer. This wave appears more intense and diffuse, particularly for smaller gaps, and follows the ionization channel created by the primary streamer. The return ionization wave demonstrates a faster propagation speed than the primary streamer. As another example, for a 6 mm-$L_{Gap}$, the counter-propagating wave travels at approximately 250 km/s, while the primary streamer moves at around 75 km/s (**Figure 6c**).
- *Secondary Streamer Propagation*: Once the reversed ionization wave reaches the tip of the transmission line, a secondary streamer propagates toward the target along the previously ionized path. This secondary streamer appears more diffuse, with a uniform luminous intensity along its channel (mostly seen in **Figure 6b**).
- *Secondary Streamer-Target Interaction*: As the secondary streamer contacts the target, its intensity is enhanced by residual charges from the primary streamer, resulting in a highly luminous interaction (mostly seen in **Figure 6a**).
- *Secondary Streamer Extinction*: Unlike the primary streamer, which initiates an immediate counter-propagation upon

contact with the target, the secondary streamer remains in contact with the target until it extinguishes (**Figure 6a**).

For gap distances higher than 10 mm (**Figures 6f–6i**), the streamer dynamics differ significantly:
- *Primary Streamer Propagation*: In contrast to smaller gaps, the primary streamer's intensity is maximum at the tip of the transmission line, then decreases during propagation and decelerates as it approaches the target (mostly seen in **Figure 6g**).
- *Primary Streamer-Target Interaction*: As in the smaller gaps, a highly intense contact point forms upon the streamer's arrival at the target, lasting for a few nanoseconds (mostly seen in **Figure 6f**).*No Counter-Propagation*: The loss of the streamer's head intensity during propagation and its slow down near the target, prevents the formation of a secondary counter-propagation wave (mostly seen in **Figure 6h**). This behavior is attributed to increased recombination processes, which are facilitated by the weaker electric field and the higher proportion of air molecules at greater distances from the catheter extremity.

For $L_{Gap}$ = 18 mm (**Figure 6i**), no interaction is observed between the guided streamer and the target. This scenario may result from the loss of the electrical charges carried by the streamer before reaching the target (around 16 mm), similar to the behavior in free-jet conditions.

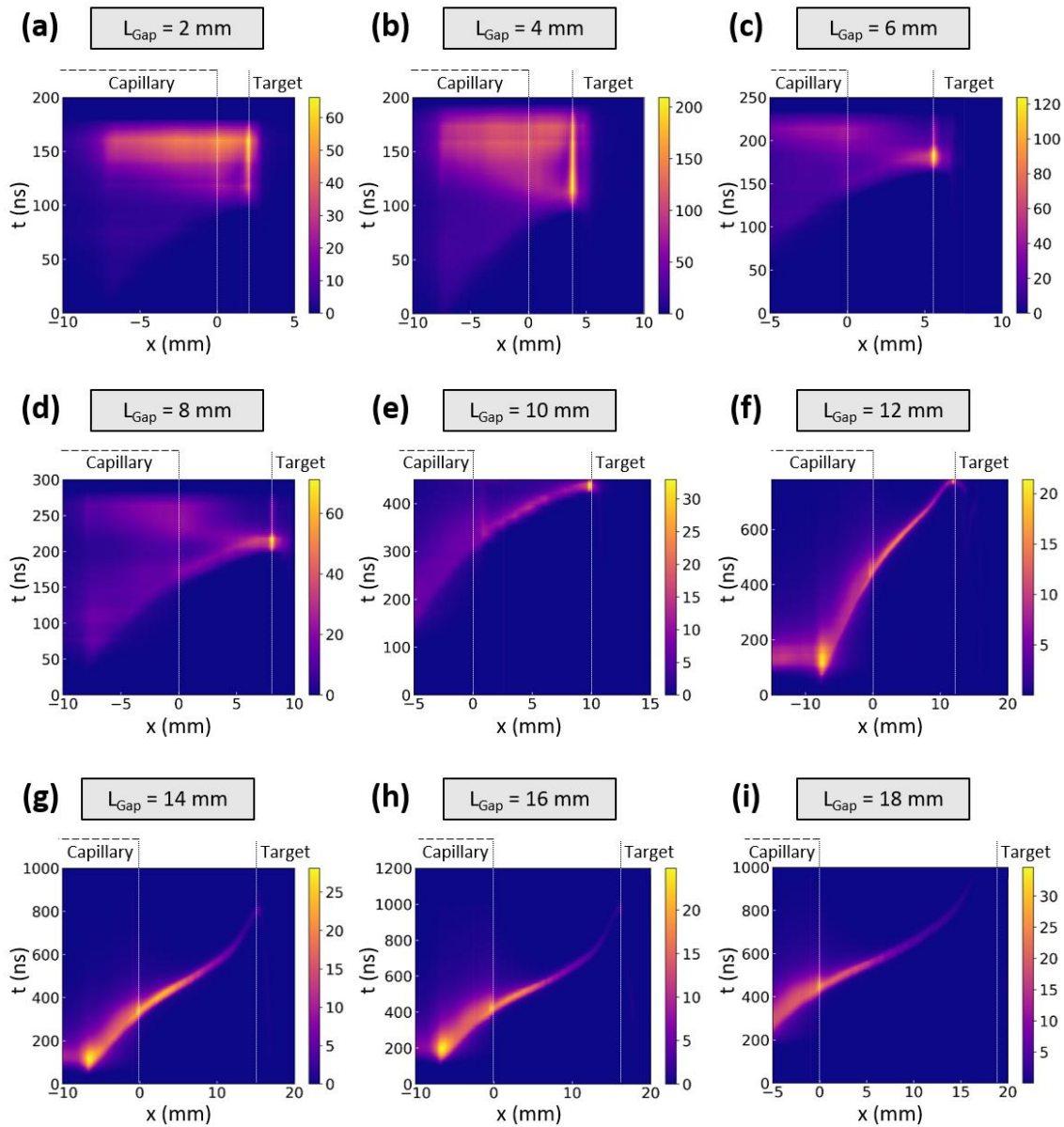

*Figure 6. Spatio-temporal representation of the streamer propagation for different gap distances. (a to i) $L_{Gap}$ is ranging from 2 to 18 mm with a 2 mm-step.*

The discrepancies in streamer propagation observed in **Figure 6** for gap distances below and above 12 mm are further analyzed in **Figure 7a,** which plots the streamer's head location ($x_{sh}$) versus time for selected gap distances. Across all gap distances, the primary streamer initiates at the capillary outlet ($x_0$) and reaches the target ($x_T$) after progressively longer times as $L_{Gap}$ increases. This behavior aligns with the increasing distance the charges must travel to reach the conductive target. The propagation kinetics of the primary streamers in the gap region follow a Boltzmann sigmoid law ($r^2>0.98$), particularly evidenced for high-$L_{Gap}$ cases. For low-$L_{Gap}$ scenarios, the counter-propagating streamers also exhibit a Boltzmann sigmoid kinetics ($r^2>0.94$) but propagate at higher velocities compared to the primary streamer. This increased velocity can be inferred from the steeper slopes in **Figure 7a** and is attributed to the counter-propagating streamers traveling through the pre-ionized channel created by the primary streamer.

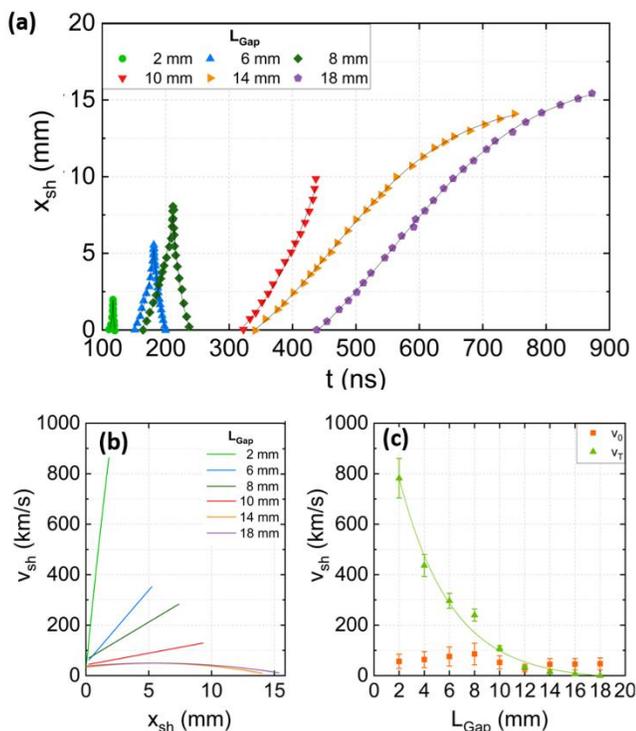

*Figure 7. Streamer location and velocity characteristics for a few gap distances. (a) Streamer head location ($x_{sh}$) as a function of time from the outlet of the capillary ($x_0$) to the target ($x_T$). The data are collected manually in the pictures of Figure 7 and the line corresponds to fits with Boltzmann sigmoid functions. (b) Instantaneous velocity of the primary streamer head versus its location along the gap. (c) Velocity of the primary streamer head as a function of $L_{Gap}$ calculated as the slope of linear regressions at the capillary outlet ($v_0$) and at the target vicinity ($v_T$).*

Additionally, the instantaneous velocity of the primary streamer is correlated to its location in **Figure 7b**. For low-$L_{Gap}$ values, a linear correlation is observed, with a decrease of the slope as the gap distance increases. In contrast, for high-$L_{Gap}$ values, the correlation becomes non-linear. The data suggest a slight acceleration of the streamer during the first 10 mm of its propagation within the gap, followed by a deceleration. This deceleration is attributed to reduced or depleted charge availability as the streamer approaches a target which is too distant to sustain propagation effectively.

The evolution of the streamer head velocity as a function of the gap distance, depicted in **Figure 7c**, is derived from linear regressions of the streamer head positions as a function of time (**Figure 7a**) at two key locations: the capillary outlet ($v_0 = v(x_0)$) and the target vicinity ($v_T = v(x_T)$). At the capillary outlet, the velocity remains consistent across all gap distances, reflecting a similar charge density at the TPC outlet. However, near the target, the streamer head velocity decreases exponentially as $L_{Gap}$ increases, aligning with the extended distance the charges must travel. The difference between the two velocities for a fixed gap ($v_T - V_0$) also tends to highlight a shift in primary streamer propagation at 12 mm-$L_{Gap}$. For low-$L_{Gap}$ values, the target velocity exceeds the one at the capillary outlet ($v_T >V_0$) while for high-$L_{Gap}$ values, the velocities converge, with a tendency of a higher velocity at the capillary outlet ($v_0>V_T$) as the target becomes more distant. These observations suggest that the effective gap distance for endoscopic procedures using this catheter should be limited to less than 12 mm to ensure successful streamer-target interaction.

### 3.4. Optical emission spectroscopy

The spectrum in **Figure 8a** is provided as a representative example to identify the reactive oxygen and nitrogen species (RONS) emitted in the plasma phase with their corresponding transitions listed in **Table 1**. More precisely, this spectrum corresponds to the emissive species produced in the capillary (x = -10 mm) and radially collected by OES for $L_{Gap}$ = 10 mm. Inside the capillary, the plasma is a mixture of helium and air, leading to high content of helium excited state (He*), observed through its atomic lines at 587.6 nm and 706.5 nm. Air compounds are mostly dominated by molecular nitrogen ($N_2$*) and nitronium ions ($N_2^{+}$*). First and second orders of the second positive system of $N_2$* are respectively headed at 337.1 nm and 674.3 nm, and $N_2^{+}$* is observed by the head of its first negative system at 391.4 nm. Moreover, the most common oxygenated radicals are identified such as hydroxyl radicals (OH*), headed at 308.9 nm and atomic oxygen (O*) at 777.4 nm. Traces of hydrogen are also collected at 656.3 nm, corresponding to the $H_\alpha$ transition.

These species are present in varying proportions throughout the plasma phase, as shown in **Figure 8b**, which shows measurements inside the capillary (x = -10 mm), just at its outlet (x = 1 mm) and at the vicinity of the metal plate (x = 9 mm). Unsurprisingly, helium is more intense in the capillary than in the plasma plume, as well as nitronium ions which are mainly produced by Penning ionization. In contrast, nitrogen molecules become predominant in the gap and far away in ambient air. Oxygen and hydrogen species are more dissociated in the capillary ($H_\alpha$ and O) due to the higher energy of electrons, while OH concentrations increase at the target vicinity.

**Figure 8c and 8d** correspond to the comparison between a gap distance of 6 mm and the previous one of 10 mm, demonstrating that smaller gaps result in more intense RONS emissions. This observation aligns with the increased streamer current (as shown in **Figure 4a**), which facilitates greater dissociation and ionization of species due to the higher charge transport and energy deposition in the plasma phase.

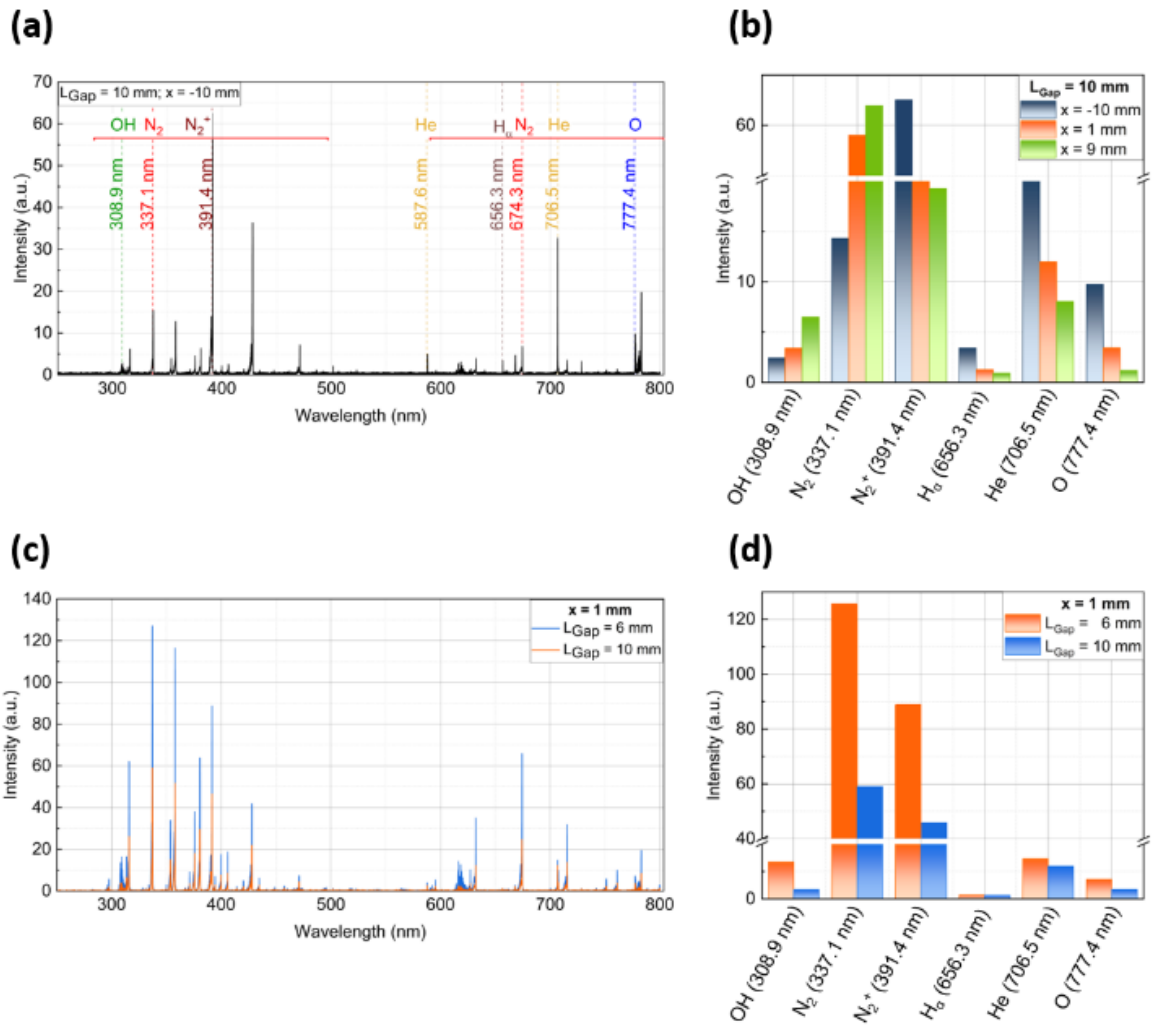

*Figure 8. Optical emission spectroscopy of the plasma phase. (a) Full spectrum from 250 nm to 800 nm, collected radially to the capillary (x = -10 mm) for a 10 mm gap. (b) Intensity of the species at the three main locations along the plasma phase. (c-d) Full spectra and main species intensity for gaps of 6 mm and 10 mm.*

| Species | λ (nm) | Lower level | Upper level | System | Ref. |
|---|---|---|---|---|---|
| OH* | 308.9 (head) | $A^2\Sigma^+$ | $X^2\Pi$ | Ang-strom | [Pearse, 1976] |
| $N_2$* | 337.1 (head) 674.3 (2nd order) | $C^3\Pi_u$ | $B^3\Pi_g$ | SPS | [Pearse, 1976] |
| $N_2^+$ | 391.4 (head) | $B^2\Sigma^+_u$ | $X^2\Sigma^+_g$ | FNS | [Pearse, 1976] |
| He I | 587.6 | 1s2p ($^3P°$) | 1s3d ($^3D$) | - | [Nist, 2024] |
| He I | 706.5 | 1s2p ($^3P°$) | 1s3s ($^3S$) | - | [Nist, 2024] |
| $H_\alpha$ | 656.3 | 2s ($^2S$) | 3p ($^2P°$) | - | [Nist, 2024] |
| O I | 777.4 | $2s^22p^3(^4S°)3s$ ($^5S°$) | $2s^22p^3(^4S°)3p$ ($^5P$) | - | [Nist, 2024] |

*Table 1. Identification of the species observed by OES in the plasma phase.*

The presence of these RONS in the low-$L_{Gap}$ values is relevant for the endotherapeutic application of this TPC because they are involved in cell death mechanisms [Privat-Maldonado, 2019], [Wende, 2018].

## 4. Conclusion

This study highlights the critical role of gap distance in the dynamics of guided streamers generated by a transferred plasma catheter (TPC) for endoscopic applications. For gap distances below 12 mm, streamers maintain their charge, successfully interact with the target, and exhibit efficient plasma propagation. Beyond this threshold, particularly at distances of 18 mm or more, the streamers lose their charge before reaching the target, rendering them ineffective. These findings emphasize the importance of optimizing gap distances to ensure effective plasma delivery during endoscopic procedures.

Moreover, this study comforts the complementarity of electric and optical measurements for a comprehensive understanding of streamer dynamics. While ICCD camera imaging effectively captures streamer propagation across the entire gap distance, its correlation with electric measurements enhances interpretation and minimizes artefacts from reflections. While the current probe proves highly effective for low gap distances, it remains undecryptable for the higher ones. Conversely, the Pockels probe, which cannot be used for smaller gap distances due to spatial constraints, provides valuable insights for larger gaps by confirming the presence of local electric fields, thus further supporting ICCD imaging data.

The results also demonstrate the robustness of the TPC design in a penalizing scenario, where the target is a grounded metal plate. This scenario provides a challenging benchmark for plasma performance, revealing significant charge dissipation and recombination processes in larger gaps due to interactions with ambient air. Despite these challenges, the TPC remains effective within clinically relevant gap distances, with charge transfer rates and current magnitudes falling within safe operational limits.

Future studies should extend these findings to clinically representative environments, including tissue-like models, to better mimic real-world impedance conditions. Additionally, exploring alternative gas mixtures and enhanced catheter designs could further improve streamer propagation, stability, and therapeutic outcomes for plasma-assisted medical applications.

# 5. Acknowledgements

——This research was supported by funding from the HRHG-CMP 2022 program of the French National Cancer Institute (Institut National du Cancer). The authors express their gratitude for the financial assistance that enabled the completion of this study. Additionally, the authors acknowledge Sorbonne Université and the Labex Plasapar for their co-funding of equipment from the Abiomede platform, which was instrumental in carrying out this work.

# 6. Conflict of interest statement

The authors have no conflict to disclose.

# 7. Data availability statements

The data that support the findings of this study are available from the corresponding author upon reasonable request.